\documentclass[iop]{emulateapj}

\def\sqig{$\sim$}

\tighten

\def\s22{1E\thinspace 2259$+$586}

\begin{document}

\submitted{}
\accepted{{\bf 1997} Journal of the British Interplanetary Society, Vol. 50, No. 7, p. 253 - 257}

\title{SETI at X-ray Energies - Parasitic Searches from Astrophysical Observations}

\author{Robin H. D. Corbet\altaffilmark{1}}
\affil{
Laboratory for High Energy Astrophysics,\\
Code 662, NASA/Goddard Space Flight Center, Greenbelt, MD 20771, U.S.A.}

\altaffiltext{1}{Universities Space Research Association}

\begin{abstract}
If a sufficiently advanced civilization can either modulate
the emission from an X-ray binary, or make use of the natural high
luminosity to power an artificial transmitter, these can serve as good
beacons for interstellar communication without involving excessive
energy costs to the broadcasting civilization. In addition, the small
number of X-ray binaries in the Galaxy considerably reduces the number
of targets that must be investigated compared to searches in other
energy bands.  Low mass X-ray binaries containing neutron stars in
particular are considered as prime potential natural and artificial
beacons and high time resolution (better than 1ms) observations are
encouraged. All sky monitors provide the capability of detecting brief
powerful artificial signals from isolated neutron stars. New
capabilities of X-ray astronomy satellites developed for astrophysical
purposes are enabling SETI in new parameter regimes.  For example, the
X-ray Timing Explorer satellite provides the capability of exploring
the sub-millisecond region. Other planned X-ray astronomy satellites
should provide significantly improved spectral resolution.  While SETI
at X-ray energies is highly speculative (and rather unfashionable) by
using a parasitic approach little additional cost is involved.  The
inclusion of X-ray binaries in target lists for SETI at radio and other
wavebands is also advocated.

\end{abstract}

\section{Introduction}

While some estimates of the number of advanced civilizations in the
Galaxy are large (e.g. 10$^6$ [1]) others suggest
that if advanced civilizations exist they should be in the solar system
given the short time required for a civilization to colonize the entire
Galaxy compared to its age. Hence, as there appears to be no reliable
evidence of extra-terrestrials within the solar system, these authors
conclude that there is no such life in the entire Galaxy (``Fermi
paradox'', [2, 3]). Yet others, however, (e.g. Brin
1983 [4]) consider numerous factors that may mitigate this apparent
paradox.

The discovery of extra-solar planets [5] together
with possible fossil evidence of early life on Mars [6]
gives hope that life may be common. However, evolutionary arguments
(e.g. Mayr, 1995 [7]) suggest that, even if life is common, {\em
intelligent} life may still be rare.
 
Despite these attempts to estimate the number of advanced
civilizations that exist in the Galaxy, the only experimentally valid
way to determine whether extraterrestrial intelligent life exists is to
try and find it.  The Search for Extraterrestrial Intelligence (SETI)
has, to date, almost exclusively concentrated on narrow band radio
signals. Principal amongst these searches have been the SERENDIP series
[8], META/BETA [9], and Project
Phoenix/HRMS [10]. While radio signals are certainly
extensively used by human civilization for communication, given that we
know nothing about how an advanced Extraterrestrial Civilization (ETC)
might attempt to communicate, it is suggested here that it is too
restrictive to investigate only narrow portions of the electromagnetic
spectrum. In particular, building on the work of Fabian [11], it is
considered whether X-ray emission might profitably be used by a
``benevolent'' ETC attempting to communicate with other intelligence in
the Galaxy and how we might detect such signals if they exist.
Although a limited optical search is underway [12] this is
viewed as problematic by some as an optical beacon must be seen against
the bright optical background of a star. In addition, it is presumed
that the ETC knows about us and deliberately targets the Earth as is
required by the very narrow optical beams. SETI by various means as
well as the radio region is discussed by Lemarchand [13].

In general, signals that are searched for in SETI may either be:
(i) accidental leakage of signals not intended to be detected
by others (e.g. analogs of television broadcasts); or (ii) deliberate
beacons.  A beacon ought to be easier to detect as it is specifically
designed for that purpose. However, even if an ETC wishes to create a
beacon for us to observe there are potential problems. For example, a
bright beacon requires the expenditure of a large amount of energy.  In
addition, how will we know where to search for the beacon? Must every
one of the \sqig10$^{11}$ stars in the Galaxy be observed?  Targeted
searches use a variety of criteria to create a manageable search (e.g.
Henry et al. [14]) which may or may not be valid.  All sky surveys,
which make no such restrictions, however, necessarily suffer from much
reduced sensitivity compared to targeted searches.

\section{Signaling in X-rays with Neutron Stars and Black Holes}

The use of nuclear weapons for signaling via X-rays was considered by
Elliott [15]. A large fraction of the energy released by a nuclear
weapon is in the form of X-rays and a high-altitude explosion by a
sufficient quantity of explosives could produce an attention-attracting
pulse.  However, as discussed by Fabian [11] a far more powerful
signal can be generated by an ETC which has the technology available to
exploit the conversion of gravitational energy to radiation via a
neutron star.  Material dropped from a large (r\,$\gg$\,10 km) distance onto
a neutron star will be accelerated to a velocity of order 0.3c and the
impact of the material onto the neutron star surface can release 10\%
of the rest mass as radiation, predominantly as X-rays. The neutron
stars that would be used in this way could be either members of
interacting binary systems (X-ray binaries) or essentially isolated
objects.

Depending on the level of technology possessed by the ETC there may be
several broadly defined ways to create an X-ray signal:\\
(i) Dump mass onto the neutron star using material such as asteroids or
comets if they exist around the star.  For comparison, some models to
explain the enigmatic $\gamma$-ray bursters involve asteroids impacting
on an isolated neutron star (see e.g. Wasserman \& Salpeter [16]). \\
(ii) ``Scavenge'' material from an accretion disk in an X-ray binary
and subsequently release it onto the compact object at desired times.\\
(iii) Directly modulate the existing mass transfer process in an X-ray
binary.  These systems are naturally variable on many time scales and it
is hence conceivably easier to modulate energy production in an X-ray
binary than, say, a star. In addition, unlike a star where photons must
slowly scatter their way out of a deep atmosphere, the photons released
by an X-ray binary escape essentially immediately from the system.\\
(iv) Obscuration, perhaps using an orbiting screen as suggested by
Fabian [11].\\
(v) ``Recycling'' of radiation. A simple scenario would be to pump an
X-ray laser [17] utilizing the energy output from the
gravitational energy release.

The various techniques need not be exclusive and some combination
could plausibly be used.

\subsection{X-ray Binaries as Beacons}
X-ray binaries are broadly divided into two groups: high mass X-ray
binaries which consist of a compact object (neutron star or black hole)
accreting material by one of several mechanisms from an early-type
companion (an O or B type star); and low mass X-ray binaries which are
compact objects accreting, typically by Roche lobe overflow, from a G
type or later stellar companion. In many cases material does not move
directly from the companion to the compact object but, due to angular
momentum, goes through an accretion disk. System sizes are typically a
few hundred light seconds for high mass systems and a few light seconds
for the low mass types.  If we assume that a sufficiently
technologically advanced civilization can manipulate X-ray emission
from an X-ray binary in some way then these sources are potential
beacons.

Attractive features of an X-ray binary as a beacon include:\\
(i) There are only a small number of X-ray binaries in the Galaxy; for
example approximately 100 bright Galactic X-ray sources are listed by
Bradt \& McClintock [18]. While there is a larger population of
transient sources the number of bright persistent sources is small. \\
(ii) They are very luminous - up to \sqig10$^{38}$ ergs s$^{-1}$.  This
has enabled some individual sources in the local group of galaxies to
be studied (e.g. Mitsuda et al. [19]).\\
(iii) High energy X-rays are not readily absorbed by the interstellar
medium unlike the case, for example, of an optical signal.\\
(iv) X-rays do not suffer from dispersion as would a radio signal.
This means that a broad-band pulse can readily be transmitted without
an observer having to search many values in ``dispersion space.''\\
(v) Accretion onto a compact object is an extremely effective way to
covert rest-mass to energy and has an efficiency of \sqig 10\% (e.g.
\"Ogelman [20]) compared to fusion (\sqig 0.1\%).  In addition,
gravitational energy can be extracted from material of any chemical
composition.\\
(vi) If X-ray sources are also generally regarded as ``interesting'' by
scientifically curious civilizations
then such a beacon is likely to be extensively observed.

If the compact object is a neutron star then releasing material onto
this object can result in the prompt production of X-rays. However, in
the case of a black hole there is no solid surface and so material can
only release energy via an accretion disk.

\subsection{Isolated Neutron Stars}

While isolated neutron stars could also be used as a beacon if an ETC
deliberately dumps mass onto it, this technique might have
disadvantages compared to exploiting natural X-ray binaries:

\noindent
(i) The neutron star is not naturally a highly luminous X-ray emitter
for a long period and hence is less likely
to be observed. Although
young short period isolated pulsars are luminous they are comparatively
short lived. There is also a much larger number of isolated
neutron stars than luminous X-ray binaries in the Galaxy (maybe
10$^8$--10$^9$ e.g.
Paczy\'nski 1990 [21], Blaes \& Madau [22])
which gives the same ``needle in a haystack'' problem as
standard radio SETI.\\
(ii) There is less likely to be an extensive supply of ``fuel'' to dump
onto the neutron star. The options of accretion disk material
extraction or direct modulation are not available.\\
(ii) The neutron star does not typically provide the natural copious
energy generation that might be of intrinsic value. The energy
output of an X-ray binary can be more than 10$^4$ times the total luminosity
of the Sun and the ETC would hence lie between type II and type
III in the classification scheme of Kardashev [23]. 

An advantage of isolated neutron stars, however, is that there are far
more of these in the Galaxy than X-ray binaries which makes them far
more accessible to an ETC.  At one point it was popular to consider
that $\gamma$-ray bursters originated from isolated neutron stars. This
would have made these candidate SETI beacons under the models discussed
here. However, the isotropic source distribution found by the Compton
Gamma Ray Observatory (CGRO) (e.g. Briggs [24]) has caused many to
abandon the single neutron star hypothesis and instead place these
objects at cosmological distances and to invoke more energetic scenarios
such as coalescing neutron star binaries (see e.g. Lipunov et al.
[25]).

If we are to find intermittent emission from an isolated neutron star
then some type of ``all sky'' X-ray detector is required.  The
properties of a number of past, present and planned experiments are
listed in Table 1. Detailed observations of a new source can typically
be done with greater sensitivity with a pointed experiment once it has
been discovered with an all sky detector. All Sky Monitors (ASMs)
typically consist of large field of view instruments with moderate
spatial resolution that scan a large fraction of the sky with some duty
cycle.

\subsection{Creating a Beacon}

This use of a natural beacon that is not located at
the ETC's ``home'' requires that it
is sufficiently benevolent and/or foresighted to send a probe to an X-ray source. If travel speed is sub-luminal this places
constraints on the life times of natural beacons that can be exploited.
For example, if a probe travels at
0.1\%$c$ to reach an X-ray binary which is perhaps located
10,000 light years away, then the natural source must have a lifetime
of at least \sqig10$^{7}$ years to make this worthwhile.
High mass X-ray binaries with lifetimes of \sqig10$^6$ years (e.g.
Verbunt [26]) would thus perhaps be less likely to be good targets.
However, low mass X-ray binaries, which may have ages in excess
of 10$^9$ years, could make much better candidates. The unusual
system Hercules X-1,
which possesses some properties of both high and low mass
systems (see e.g. Bradt \& McClintock [18]), is also thought to be relatively old and is thus
also a candidate. For a journey
of this distance some type of self-replicating (``von Neumann'')
probe may be required [27]. If a long travel
time is indeed required then, to make this worthwhile, the beacon
itself might also be expected to have a very long lifetime.
The foresight to create such a beacon is on a grander scale than, but
perhaps philosophically comparable to, the plaques which have been
placed on the Voyager 1 and 2, and Pioneer 10 and 11 probes. The long
lifetimes of low mass X-ray binaries might also be attractive to a
civilization which desires some type of enduring monument to its
existence (compare, for example, to ancient Egyptian pyramids).  The
low mass X-ray binaries also exhibit a wide range of, apparently
natural, variability such as quasi-periodic oscillations and bursts
which are causing their temporal variability to be studied in detail.
Restricting targets to low mass neutron star X-ray binaries would
reduce still further the number of potential beacons to be
investigated. However, this would certainly be too restrictive given
the small gain in efficiency compared to investigating all X-ray
binaries.

If the ETC is only capable of relatively limited low intensity
modulation of X-ray emission it may choose to do so at times when the
``normal'' emission is at relatively low levels in these naturally
variable systems.  Conversely, if the ETC is capable of triggering
transient sources to go into outburst then large outbursts from
binaries should be observed in detail (as they are likely to be
anyway).  A hint that such an amplification mechanism might be possible
comes from accretion disks instability models of transients which show
that a state change in a small region of a disk can rapidly spread to
the rest of the disk (e.g.  Cannizzo [28]) and hence a small change
could be amplified.  Another situation where a small change in the
input results in a large change in the output is where mass accretion
rate is just below the value required to overcome the centrifugal
barrier to accretion caused by the strong magnetic field of a rotating
neutron star [29].

\section{Types of Signal}

A beacon might consist of three components:\\
Type - I: A ``look at me" very strong signal of low or zero
artificial signal content.\\
Type - II: A ``carrier" signal indicating we should look further.\\
Type - III: A high information capacity signal.

For an X-ray binary these three types of signal might correspond
to:\\
(i) The natural variable X-ray emission.\\
(ii) Some type of artificial signal indicating the presence of
intelligence. For example this might be the first hundred digits of the
number $\pi$ pulse encoded in binary.  If only a comparatively weak
signal can be generated then it would be especially advantageous to
repeat this in a highly periodic way to increase the possibility of
detection.\\
(iii) The high information content signal might either be
in the X-ray band or could be in another wave band, such as radio,
where it might be easier to create a high telemetry rate signal.
If in the X-ray band, the signal might be less blatant than the
Type - II signal, for example it might utilize one or more narrow
energy bands. For example, using the ``recycling'' technique.

\section{Disadvantages of X-rays}
Communication via X-rays does, naturally, suffer from limitations.  For
example, individual X-ray photons have a large amount of energy and
communication rates are hence set by the number of photons that can be
detected rather than the frequency of the X-rays.  In addition
the mass-dumping technique does not create a narrow energy band signal, unlike
most radio or optical transmission techniques, which reduces the
signal-to-noise level.
However, neither of these are severe problems,
particularly for Type - I and Type - II signals where it might be
preferable to send a broad band signal when sufficient energy is
available to ensure that it is detected.

An additional difficulty with X-ray observations is that
they must be conducted from space. While, in itself, this is
not a major problem, it does typically result in a restriction
on data telemetry rates compared to an entirely ground-based system.
High telemetry rates will be produced when both high spectral
and high temporal resolutions are required.
This problem could be alleviated by
searches for signals on board the spacecraft by a sufficiently
powerful onboard computer - the raw
data itself is not telemetered. An extension of this would be
to have a high data rate connection to a space station where
the data are stored and/or processed.
%

\section{High Time Resolution X-ray Observations}

If significant artificial modulation of emission from X-ray binaries
exists why has it not yet been seen? Fabian [11] claimed that signals
that could be seen with simple detectors could relatively easily be
produced.  While astrophysical X-ray data are not typically
investigated for the presence of artificial signals, they are often
subjected to a variety of timing analyses such as performing Fourier
Transforms which can reveal modulation such as a periodic signal.
Although we assume that the ETC is benevolent, an entity that has
sufficient foresight to create a very long lasting beacon 
may not necessarily create a signal that can be detected with the
crudest X-ray detectors: X-ray astronomy is a young subject and
extra-solar X-ray sources have only
been studied since 1962 [30]. The signal may simply be too weak for us
to have detected so far; we cannot determine {\sl a priori} how much of
the emission from an X-ray binary might be modulated by a highly
advanced civilization.  However, an intriguing possibility may be that
we are about to open new ranges of parameter space which could contain
some type of signal.  This is based on the possibility that Type II or
III signals may be modulated comparatively rapidly and/or utilize a
narrow energy band.

The Rossi X-ray Timing Explorer (RXTE; [31]) was
launched in late 1995. While the main Proportional Counter Array (PCA)
detector will offer several improvements over previous satellites in
terms of a large collecting area (~7000cm$^2$) and a reasonable
telemetry rate (mean of \sqig40 kbps with a maximum rate of 512 kbps
for short periods) the significant new regime of parameter space that
will be opened is for very rapid variability.  The PCA consists of five
separate proportional counters; individual X-ray photons can be
time-tagged to 1$\mu$s accuracy and each counter has an independent
dead time of approximately 10 $\mu$s.  XTE hence offers the potential
to explore sub-millisecond down to microsecond timing not readily
accessible to previous X-ray satellites. One driving force behind
including this high time resolution capability is to study rapid
variability in black hole X-ray binaries such as Cygnus X-1 [32]. 
The RXTE PCA is compared to some other X-ray missions in
Table 2.

Rapid modulation might be employed by an ETC for several reasons. For
example, the more rapid the variability, the greater the telemetry rate
that results, this could be valuable for Type - II or III signals.
Rapid variability may also be advantageous if, on these timescales,
there is much less natural signal. For example the minimum rotation
period of a neutron star is of order 1 millisecond as is the period of
a Keplerian orbit at a neutron star surface.  Observations of many
systems show the presence of low-frequency timing noise which again
implies rapid modulation could be advantageous.  Natural quasi-periodic
signals at frequencies up to at least ~kHz frequencies have been
detected [33] suggesting artificial modulation
frequencies at least this high should be used.

Theoretical investigations of natural emission from blobs falling on
neutron stars show X-rays can be released which are modulated on
microsecond timescales [34]. Hence, there appears to
be no compelling reason that artificial modulation on the same
microsecond time scale could not also be produced.  A Type - II signal
might well be periodic on some time scale and could perhaps be detected
as a byproduct of searches that are made to detect the rotation periods
of neutron stars in low mass X-ray binaries.

\section{High Spectral Resolution X-ray Observations}
Currently being planned are at least two X-ray satellites which will
provide a combination of high spectral resolution, potentially high
temporal resolution, and reasonable collecting areas. These are the
Japan/US mission Astro-E, which is scheduled for launch in 1999, and
NASA's HTXS (High Throughput X-ray Spectrometer) which is currently in
a design phase. The drive for higher spectral resolution comes from a
desire to study in detail emission lines such as those from iron in the
6 -- 7 keV range as well as others at lower energies. These
significantly improved energy resolutions are being achieved by
exploiting devices such as
micro-calorimeters (e.g. Moseley et al.
[35]) and superconducting tunnel junctions as
X-ray detectors [36, 37].

If the ETC is capable of more sophisticated modulation of X-rays, such
as confining a signal to a narrow energy band by, for example, using an
X-ray laser this will also make it less likely that we would have
detected such a signal especially given the very limited energy
resolution of current X-ray detectors.  Proportional counters, for
example, have resolutions of \sqig20\% at 7 keV while CCD detectors
such as those on ASCA, which have a factor \sqig10 better energy
resolution, are often used with time resolutions of seconds [38].
These new detectors thus have the potential to
significantly expand still further the range of parameter space through
which X-ray SETI can be performed.

\section{Conclusion}
The  development of new detectors continues to expand the parameter
space over which we can perform X-ray SETI. Further, in a somewhat
similar philosophy to that adopted by SERENDIP [8] in
that SETI does not affect regular astrophysical observations, all high
time and/or spectral resolution X-ray observations of X-ray binaries
can be investigated for unusual signals. The advantage over SERENDIP is
that not even any additional hardware is required. Rapid periodic
signals during short ``bursts'' during otherwise low level emission
from an X-ray binary could be candidate signals from an ETC.
Alternatively, if the ETC can trigger transient outbursts then high
time resolution observations of these should be made. There are already
in place several ``Target of Opportunity'' programs to undertake such
observations of  transients that are detected with the All Sky Monitor
experiment on XTE.  Other X-ray astronomy satellites that are being
planned will also offer significantly enhanced spectral resolution
which will open yet another parameter regime.  As data from current and
future X-ray missions enter public archives it may be profitable to
perform systematic searches on all data for pulsed ``narrow'' energy
band signals. While a coherent pulsed signal does not, on its own,
guarantee that artificial modulation is present, such signals,
especially those with periods less than the minimum rotation period of
a neutron star, should be investigated in greater detail.

Including bright X-ray sources in radio and other targeted searches may
also be worthwhile. For example, the ETC might construct a higher
information capacity optical or radio beacon that will be found once
the observer has been alerted to the artificial signals arising from
the X-ray source. This Type - III information beacon could make use of
the copious long lived energy production from the X-ray binary as its
power source.

Given our ignorance of how advanced ETCs might attempt to communicate
with us, restricting searches to just the radio portion of the
electromagnetic spectrum is likely to be far too limiting.  It was once
proposed that forests be planted in geometric shapes to communicate to
extra-terrestrials that intelligent life exists on the Earth. In just a
few hundred years time will it still seem as obvious that radio waves
are the best means to communicate across interstellar or intergalactic
distances? While it is not claimed here that X-rays are necessarily
used by advanced ETCs for communication, it is emphasized that all
potential communication channels should be investigated.  As X-ray
detector technology improves, in addition to revealing more about the
natural high-energy Universe, we are also providing additional channels
where evidence of an ETC might be found.

The scenario discussed here and as originally proposed by Fabian
[11], while highly speculative, does not require the existence of
numerous long-lived civilizations in the Galaxy, neither is it
necessarily required that a large fraction of the Galaxy has been
colonized. What is required, however, is that, at some time in the
history of the Galaxy a civilization existed
with the desire and technological capability to create a durable
beacon. The factor ``L'' in the Drake equation is not the lifetime of
the civilization itself but that of whatever beacon it can create.

One way around the Fermi paradox is the proposal that ETs will not be
found everywhere in the Galaxy. Instead they will only be in the
``interesting'' places (e.g. Shostak [39]). While the Solar System
itself might not be regarded as interesting, the comparatively rare
luminous neutron star and black hole binary systems may be more
attractive places.

\newpage
\clearpage

\begin{table}
\caption{Selected X-ray Astronomy Satellites - All Sky Monitors}
\begin{center}
\begin{tabular}{lrrrc}
Satellite/Instrument & Band Pass& Angular Resolution& Sensitivity &
Mission Dates\\
          & (keV) & (degrees)& ($\mu$Jy)& \\
\tableline
Vela 5B (XC) & 3-12 & 6.1 & 400 & 1969 - 1979\\
Ariel V (ASM) & 3-6 & 10 & 170 & 1974 - 1980\\
Ginga (ASM) & 2-20 & 0.2 & 50 & 1987 - 1991\\
Granat (Watch) & 6-180 & 2 & 100 & 1989 - ?\\
CGRO (BATSE)$^1$ & 20-600 & 5 & $(a)$ & 1991 - ?\\
XTE (ASM) & 2-10 & 6 & 25 & 1995 - ?\\
SAX (WFC)$^2$ & 2-30 & 0.1 & 1 & 1996 - ?\\
Spectrum-X/$\gamma$ (MOXE) & 2-25 & 1.1 & 7 & 1997? - ?\\
\tableline
\end{tabular}
\end{center}
Notes: taken in part from In't Zand, Priedhorsky \& Moss [40],
and references therein. Also, (1) Horack [41], (2) Jager et al. [42].
(a): 3$\times$10$^{-8}$ ergs cm$^{-2}$ for a one second burst.
Angular resolutions are worst cases for a particular instrument.
\end{table}

\begin{table}
\caption{Selected X-ray Astronomy Satellites - Pointed Experiments}
\begin{center}
\begin{tabular}{lrrrc}
Satellite/Instrument & Time Resolution& Spectral Resolution& Collecting Area &
Mission Dates\\
          & ($\mu$s)               & $E/\Delta E$& cm$^2$& \\
\tableline
Ginga (LAC) & 980 & 6 & 4000 & 1987 - 1991\\
ASCA (SIS) & 16,000 & 50 & 250$\times$2 & 1993 - ?\\
XTE (PCA) & 1 & 6 & 7000 & 1995 - ?\\
Astro E (XRS) & 20? & 670 & 400 & 2000? - ?\\
HTXS & ? & \sqig10000? & 2000? & $>$2000? - ?\\
\tableline
\end{tabular}
\end{center}
Notes: Spectral resolutions are approximate and are at 6.7 keV. Time
resolutions are the best available for an instrument and are not necessarily
generally used. Extensive information on X-ray astronomy missions
can be found in Bradt, Ohashi \& Pounds [43].
\end{table}

\end{document}